%% file: ijcai25.tex
\documentclass{article}
\usepackage{ijcai25}

\usepackage{times}
\usepackage{soul}
\usepackage{url}
\usepackage[hidelinks]{hyperref}
\usepackage[utf8]{inputenc}
\usepackage[small]{caption}
\usepackage{graphicx}
\usepackage{natbib}
\usepackage{amssymb}
\usepackage{amsmath}
\usepackage{amsthm}
\usepackage{booktabs}
\usepackage{algorithm}
\usepackage{algorithmic}
\usepackage[switch]{lineno}
\usepackage{balance}

\usepackage{xcolor}

\title{Generative Augmented Reality: Paradigms, Technologies, and Future Applications}

\author{
Chen Liang$^{1,4}$ \and
Jiawen Zheng$^1$ \and
Yufeng Zeng$^1$ \and
Yi Tan$^3$ \and
Hengye Lyu$^1$ \and
Yuhui Zheng$^1$ \and \\
Zisu Li$^2$ \and
Yueting Weng$^4$ \and
Jiaxin Shi$^4$\And
Hanwang Zhang$^3$
\\
\affiliations
$^1$The Hong Kong University of Science and Technology (Guangzhou)\\
$^2$The Hong Kong University of Science and Technology\\
$^3$Nanyang Technological University $^4$XMax.AI Ltd.\\
\emails
Contact: chenliang2@hkust-gz.edu.cn, jiaxin@xmax.ai, hanwangzhang@ntu.edu.sg
}

\begin{document}

\maketitle
\begin{figure*}[t]
  \centering
  \includegraphics[width=\textwidth]{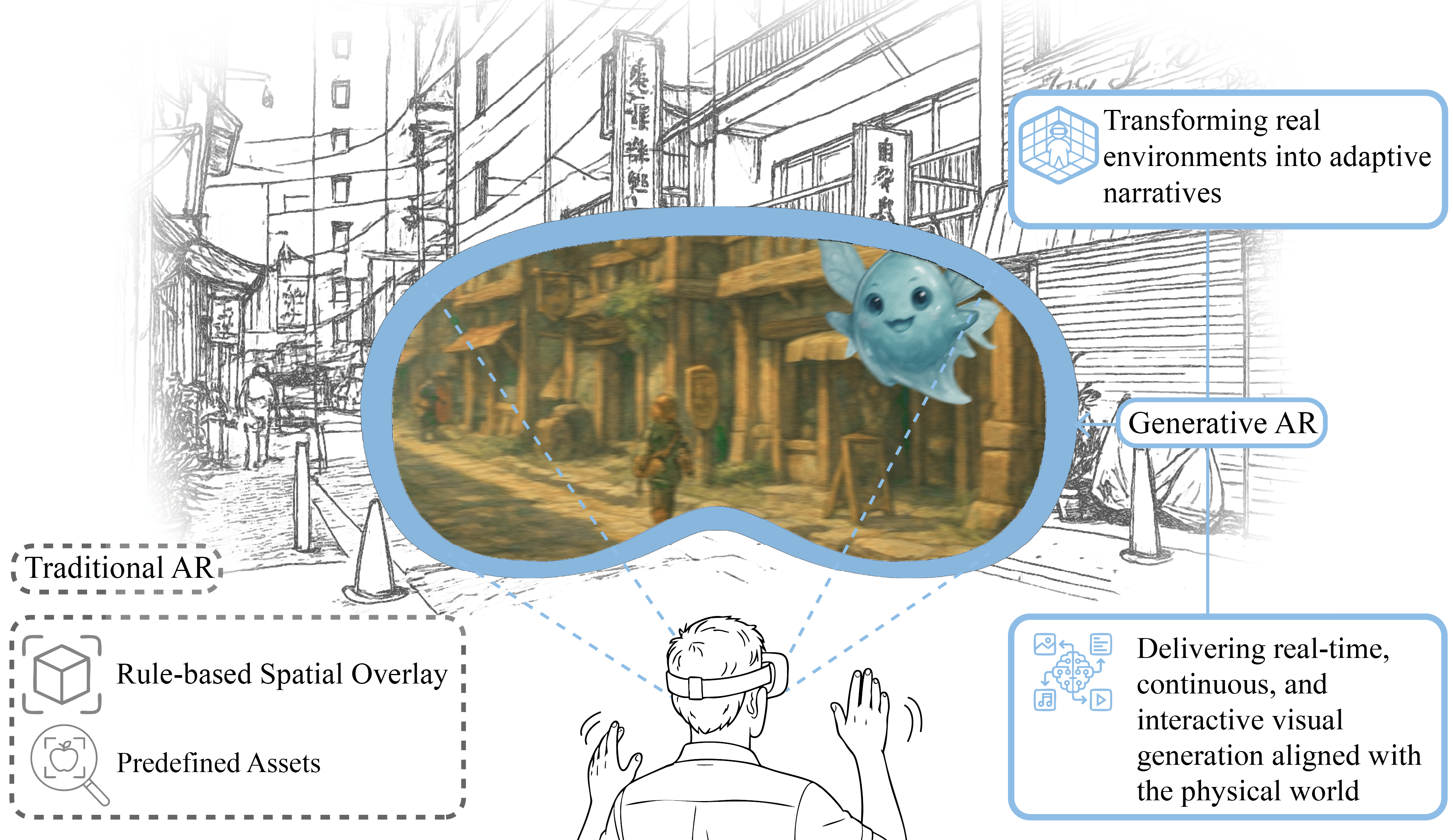}
  \caption{Conceptual illustration of Generative Augmented Reality (GAR). Traditional AR relies on rule-based spatial overlays and predefined assets, whereas GAR integrates environmental cues into a unified generative process that re-synthesizes the visual scene in real time. This enables adaptive narratives and continuous, interactive visual augmentation aligned with the physical world.}
  \label{fig:teaser}
\end{figure*}

\input{secs/0_abstract}

\input{secs/1_introduction}

\input{secs/2_paradigm}

\input{secs/3_foundations}

% ...

\input{secs/4_applications}
\input{secs/6_conclusion}






\bibliographystyle{named}
\bibliography{ijcai25}

\end{document}

%% file: secs/0_abstract.tex
\begin{abstract}
    This paper introduces \emph{Generative Augmented Reality} (GAR) as a next-generation paradigm that reframes augmentation as a process of \textit{world re-synthesis} rather than world composition by a conventional AR engine.
    GAR replaces the conventional AR engine’s multi-stage modules with a unified generative backbone, where environmental sensing, virtual content, and interaction signals are jointly encoded as conditioning inputs for continuous video generation.
    We formalize the computational correspondence between AR and GAR, survey the technical foundations that make real-time generative augmentation feasible, and outline prospective applications that leverage its unified inference model.
    We envision GAR as a future AR paradigm that delivers high-fidelity experiences in terms of realism, interactivity, and immersion, while eliciting new research challenges on technologies, content ecosystems, and the ethical and societal implications.
\end{abstract}


%% file: secs/1_introduction.tex
\section{Introduction}


Augmented Reality (AR) emerged as a response to the longstanding goal of blending digital content with physical environments grounded in users’ real-world perception and action. Early formulations, such as \cite{thomas1992augmented}’s work on overlaying digital instructions for aircraft assembly and \cite{milgram1994taxonomy}’s Reality–Virtuality continuum, situated AR as an intermediate blend between virtual reality and physical reality. As advances in sensing, spatial tracking, and real-time rendering \citep{azuma1997survey} made it possible to align digital content with physical scenes, AR evolved into a technology framework for enabling users to perceive and interact with virtual elements as part of their surroundings, which is widely adopted in domains such as industrial guidance, education, navigation, and interactive media.

However, as technological progress elevates expectations for content fidelity, interaction precision, and naturalistic responsiveness in AR, the compositional paradigm underlying conventional AR architectures reveals inherent constraints. Existing systems typically rely on explicitly modeled assets, predefined interaction rules, and deterministic graphics pipelines. This structure makes it difficult to synthesize high-fidelity interactions, such as fluid material behaviors, complex mechanical dynamics, and even the responsiveness of living creatures. Scaling toward broader expressive spaces often increases authoring burden and system fragility: producing high-fidelity 3D assets demands substantial manual labor, yet even meticulously crafted assets possess limited behavioral expressiveness, making it difficult to achieve truly responsive or realistic interactions.


In parallel, the rapid advancement of generative models, particularly in diffusion-based video generation models \citep{ho2022video,kong2024hunyuanvideo}, has introduced a fundamentally different way of constructing visual experience. These models are capable of producing temporally coherent, semantically grounded videos of and beyond both the physical and the imaginary world contents from high-level conditions such as textual intent \citep{luo2023videofusion}, motion cues \citep{bai2025recammaster}, reference frames \citep{hu2024animateanyone}, or behavioral signals \citep{guo2025mineworld}. Rather than treating scenes as fixed backdrops for augmentation, generative video models represent reality as a learnable, extendable process, where physical consistency and temporal evolution are expressed within a unified latent space. As such models progress toward real-time inference \citep{yin2025slow} and controllable streaming \citep{lin2025autoregressive}, they shift the computational focus from overlaying content to generating world evolution under interaction.

This paper presents a forward-looking conceptual and technical survey of Generative Augmented Reality as a computational framework for next-generation spatial computing. Our contributions are threefold:
\begin{itemize}
    \item We formalize the computational transition from compositional AR pipelines to generative world re-synthesis, providing a comparative formulation of their perceptual grounding, control flow, and asset management, and rendering mechanisms.
    \item We survey the enabling technologies underlying GAR, including streaming video generation models, computational efficiency and quality optimization, multimodal control mechanisms, and asset management.
    \item We analyze the future application landscape of GAR and its potential to reshape spatial experience, embodied creativity, adaptive storytelling, collaborative world-building, and mixed-reality ecosystems.
\end{itemize}


%% file: secs/2_paradigm.tex
\section{Generative Augmented Reality: The Next Generation of Spatial Computing and Interaction Paradigm}

In this section, we present the paradigm of Generative Augmented Reality (GAR) in the context of the rapid development of generative video models. GAR rethinks the pathways to achieve augmentation of reality, representing a shift in the technical architecture from explicit 3D modeling to implicit world video-stream re-rendering.

To ground this paradigm, we first revisit the fundamentals of traditional augmented reality (AR), outline its technology stack and implementation hierarchy, and then explain how GAR transforms this architecture into a model-driven framework that performs implicit, real-time, and generative re-rendering of the real world.




\subsection{Concept of Augmented Reality}



The conceptual basis of AR was first formalized by \citet{milgram1995augmented} through the Reality–Virtuality Continuum, which positioned AR within a spectrum ranging from purely physical to fully virtual environments.
Later, \cite{azuma1997survey} provided a widely accepted operational definition, identifying three essential characteristics of AR systems: 1) combination of real and virtual content, 2) real-time interactivity, and
3) accurate three-dimensional registration.

Building on these principles, \citet{craig2013understanding} and \citet{billinghurst2015survey} summarized AR as a multidisciplinary synthesis of computer vision, graphics, sensing, and interactiton—designed to enable spatial coherence between the physical and virtual worlds.
These frameworks define AR as a perception–action feedback loop where tracking, registration, rendering, and interaction form the core of the experience.

With recent works, \citet{mendoza2023augmented} highlight advances in semantic anchoring and adaptive context modeling that extend AR beyond geometric registration, while \citet{auda2023scoping} frame AR within cross-reality systems emphasizing embodied and context-driven interaction. Together, these recent perspectives expand foundational definitions by positioning context awareness, semantic understanding, and embodied engagement as defining attributes of next-generation AR paradigms.

\subsection{Traditional Augmented Reality Architecture and Technical Stacks}








Following canonical AR engine architectures \cite{craig2013understanding,zhou2008trends,billinghurst2015survey}, a traditional AR system can be decomposed into four tightly coupled subsystems: 1) tracking and world understanding, 2) scene management and anchoring, 3) rendering and asset handling, and 4) interaction and input control. Together, these components form the runtime loop of an AR engine, maintaining spatial consistency between the physical and virtual worlds.

\paragraph{Tracking and World Understanding}

The foundational layer of any AR engine performs environment tracking and spatial understanding.
This involves pose estimation, geometric reconstruction, lighting estimation, and semantic understanding.  
Classic monocular SLAM systems, such as those introduced by \citet{mur-artal2015orbslam} and extended by \citet{campos2021orbslam3} and \citet{qin2018vinsmono}, provide reliable localization for mobile and wearable AR.
\citet{newcombe2011kinectfusion} introduced dense mapping techniques that enhanced geometric fidelity, while \citet{li2024rdvio} proposed a visual–inertial odometry method improving robustness under fast motion.
Modern frameworks such as ARKit and ARCore now integrate plane detection \citep{kim2022integrating}, depth reconstruction \citep{ganj2023mobile,zhang2022indepth}, and lighting estimation \citep{somanath2021hdr,zhao2022litar}, combining these with semantic segmentation and object recognition \citep{oufqir2020arkit} to support context-aware anchoring and dynamic physics simulation.


\paragraph{Scene Management and Anchoring}


Coordinate systems and anchors define how virtual content aligns with the real world.  
Anchors maintain persistence and spatial stability under camera motion, bridging sensing and rendering \citep{maio2022augmented}.  
Cloud-based anchoring mechanisms, such as ARCore Cloud Anchors and ARKit geospatial anchors enable cross-session and multi-user spatial consistency \citep{oufqir2020arkit}.  
Semantic segmentation and object recognition (e.g., Mask R-CNN \citep{he2018mask}) further guide anchor placement by aligning virtual content with the physical scene’s semantics and topology \citep{dechicchis2020semantic}.

\paragraph{Rendering and Asset Pipeline}

Rendering subsystems handle the compositing of virtual elements with the camera feed.
It manages photometric consistency in lighting, shadow, and occlusion \citep{somanath2021hdr,zhao2022litar}.  
Rendering engines such as Unity and Unreal provide shader integration and runtime resource management for 3D models, textures, and animations, which are bound to anchors and displayed as overlays \citep{vidal-balea2025offcloud}.  
Recent neural rendering advances, including Neural Radiance Fields (NeRF) \citep{mildenhall2020nerf} and generalizable variants (Gen-NeRF) \citep{fu2025gennerf}, enable photorealistic view synthesis and occlusion-aware compositing, supporting higher-fidelity AR visualization.

\paragraph{Interaction Framework and Input System}

The interaction framework governs user input, event dispatching, and multimodal feedback.
It integrates diverse input modalities such as gesture, voice, gaze, and touch, which are mapped to AR events (e.g., select, drag, scale) \citep{karakostas2024realtime}.
Recent studies emphasize that multimodal fusion and context-adaptive feedback are key to achieving natural and seamless interaction in AR/VR environments \citep{dritsas2025multimodal}.
Systems combining gaze tracking, speech recognition, and spatial audio provide more intuitive and embodied interactions, thereby enhancing user immersion.
Ultimately, this layer closes the control loop by coupling perceptual augmentation with human intention.

\subsection{Generative Augmented Reality Paradigm}

A conventional augmented reality (AR) engine operates as a closed computational loop integrating \textbf{perception}, \textbf{rendering}, and \textbf{state update}. At each timestep $t$, the AR system perceives the physical world $\text{World}_t$ through on-device hardware (e.g., cameras, inertial units, microphones, and controllers) using specific recognition models, yielding environmental observations $\mathbf{s}_t^{\text{env}}$ (e.g., spatial geometries and objects) and user behavior signals $\mathbf{s}_t^{\text{int}}$ (e.g., head motion, hand pose, and gaze):
\[
(\mathbf{s}_t^{\text{env}},\, \mathbf{s}_t^{\text{int}}) = F_{\text{sense}}(\text{World}_t).
\]
These sensory observations serve as constraints of the AR engine and renderer. Combining $\mathbf{s}_t^{\text{env}}$, $\mathbf{s}_t^{\text{int}}$ with the current asset state $\mathcal{A}_t$ (e.g., the states and attributes of virtual scenes and objects), the AR engine simulates the physics and interaction effects among real-world perceptions and virtual assets, updating the virtual asset state andproducing the next augmented frame:
\[
\begin{aligned}
\mathcal{A}_{t+1} &= 
F_{\text{sim}}\!\big(\mathcal{A}_t,\, \mathbf{s}_t^{\text{int}},\, \mathbf{s}_t^{\text{int}}\big), \\
\mathbf{x}_{t+1}^{\text{AR}} &= 
F_{\text{render}}\!\big(\mathcal{A}_{t+1},\, \mathbf{s}_t^{\text{env}},\, \mathbf{s}_t^{\text{int}}\big).
\end{aligned}
\]


\textbf{Generative Augmented Reality (GAR) represents a paradigm shift from rule-based to neural-based computation}, which is analogous to the evolution from early scripted chatbots to large neural conversational models such as ChatGPT. 
Such a transition redefines the boundary of augmentation in terms of representation capability and interaction freedom. In execution, GAR retains a similar conceptual framework but collapses simulation and rendering stages into a unified generative process driven by a single model $\mathcal{G}_\theta$.
Given the sensed observations and contextual conditions, the model directly re-synthesizes the next world frame as
\[
\mathbf{x}_{t+1}^{\text{GAR}} = 
\mathcal{G}_\theta\!\big(\mathbf{x}_{\le t},\, \mathbf{s}_t^{\text{env}},\, \mathbf{s}_t^{\text{int}},\, \mathcal{C}_t\big),
\]
where $\mathcal{G}_\theta$ refers to the generative model, $\mathbf{x}_{\le t}$ denotes the previously generated frame sequence,
$(\mathbf{s}_t^{\text{env}}, \mathbf{s}_t^{\text{int}})$ are the environmental and interaction observations,
and $\mathcal{C}_t$ represents conditioning information that includes static references
(e.g., prompt embeddings, reference images, or latent asset descriptors)
as well as memory-dependent assets that evolve through long-term generative context.

Unlike conventional AR, GAR does not explicitly maintain or update virtual assets through frame-wise simulation.
Instead, the evolution of memory-related assets—such as persistent object states, scene history,
or user-specific visual traces—is implicitly handled through the model’s recurrent latent dynamics and context encoding.
Given these conditions, the model automatically performs implicit physical reasoning, visual synthesis,
and interaction response within a single generative process,
producing temporally coherent augmentation without explicit scene management or layered compositing.

This generative formulation preserves the temporal and perceptual continuity of the AR loop 
while fundamentally altering how assets, computation, and control signals are represented and processed.
By embedding physical reasoning and rendering within a unified latent process,
GAR redefines the design constraints of augmented reality systems—simplifying content authoring,
stabilizing performance, and extending the expressive range of interaction.
The following subsections outline these architectural and experiential advantages in detail.

\paragraph{Asset Simplification and Implicit Representation}
GAR eliminates the need for complex and explicitly authored scene assets.
Traditional AR pipelines depend on pre-defined meshes, materials, and textures,
whereas GAR can condition generation on a single reference image, depth map, or textual description.
This implicit representation drastically lowers the authoring barrier
and allows dynamic generation of scene-consistent content without manual modeling.

\paragraph{Computational Stability and Scalability}
In conventional AR engines, the computational load scales with scene complexity,
as physics simulation and rendering pipelines must account for object count, lighting, and occlusion.
By contrast, GAR maintains a near-constant computational footprint,
since its generative backbone $\mathcal{G}_\theta$ operates with fixed architectural complexity.
This enables stable throughput and predictable latency, even under dense or dynamic environments.

\paragraph{Unified Computational Model}
Traditional AR relies on heterogeneous subsystems—
including scene reconstruction, gesture recognition, eye tracking, pose estimation, and object rendering—
each optimized separately and communicating through explicit data exchange.
GAR replaces these modular pathways with a single generative computation stream.
All environmental and interaction signals are injected as auxiliary conditioning inputs,
while the core inference operates within one continuous generative model.
Outside the backbone, only lightweight resource management and synchronization modules are required.

\paragraph{Multimodal Extensibility}
Because conditioning occurs within the same latent generative space,
GAR naturally accommodates multimodal control inputs such as text, voice, gaze, or contextual semantics.
Unlike conventional AR systems that require separate modules for new modalities,
GAR treats all forms of input as unified conditioning vectors,
enabling flexible interaction and intuitive adaptation across sensory channels.

\paragraph{Generative Diversity and Expanded Interaction Space}
GAR unlocks generative behaviors that are difficult or impossible to achieve in traditional AR.
The model can synthesize complex dynamics and interactive assets—such as deformable materials, soft bodies,
animals, fluids, or articulated mechanisms—without explicit physical modeling or animation rigs.
This expands the expressive and experiential scope of AR beyond overlay-based augmentation,
toward an open-ended, continuously generated world stream.

%% file: secs/3_foundations.tex
\section{Technical Foundations of GAR}

%
\begin{figure*}
  \centering
  \includegraphics[width=1\textwidth]{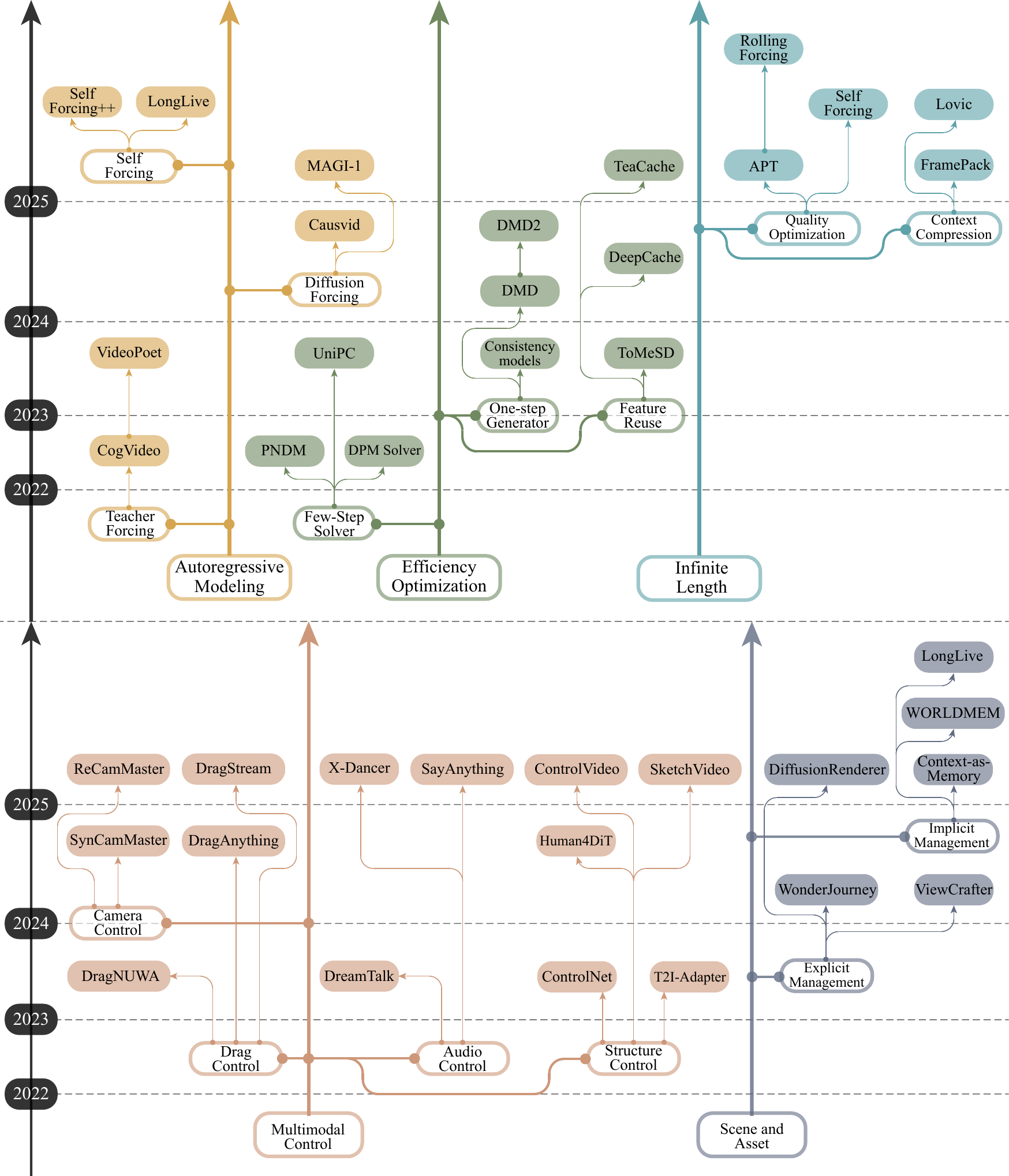}
  \caption{GAR technical foundations, mapped as an evolutionary tree from 2022 to 2025 as five branches: Autoregression, Efficiency, Infinite Length, Multimodal Control, and Scene/Asset—organize prior work and its lineage.}
  \label{fig:placeholder}
\end{figure*}

GAR is built upon advanced video generation foundation models, which endows it with the potential to generate high-fidelity video sequences with strong temporal consistency. However, GAR’s unique generation paradigm introduces two major technical challenges.
\emph{First}, 
GAR demands both high frame rates and low-latency video stream generation, which necessitates a shift in the \emph{generation paradigm} from traditional, inefficient video generation models that produce clips jointly in the spatio-temporal domain to real-time streaming video generation.
\emph{Second}, from the perspective of \emph{content quality}, 
GAR needs to continuously generate temporally consistent video sequences under diverse interactions between the physical and the digital worlds in an open-ended and dynamic environment.
This poses content quality challenges that go beyond the capabilities of current foundation video models, which are typically designed for fixed sequence lengths, single and low-interaction control, and low scene dynamics.

In this section, we first introduce the fundamental generative theories underlying video generation models in Section \ref{sec:pre}. Then, to address the \emph{first} challenge, we discuss advances and promising techniques in two directions: achieving streaming video generation through autoregressive modeling along the temporal dimension, and enhancing GAR’s efficiency by optimizing the computational load during the generation process, covered in Sections \ref{sec:ar} and \ref{sec:eff} respectively.
Following that, to tackle the \emph{second} challenge, we elaborate on three representative problems that faced by video models when generating GAR content with the potential technical solutions: infinitely long video generation and quality optimization (Section \ref{sec:long}), multimodal interactive control (Section \ref{sec:control}), and scene and asset management in video generation (Section \ref{sec:scene}).

\subsection{Preliminaries on Video Generation}
\label{sec:pre}

As the base of GAR, video generation models need to produce realistic and temporally consistent video sequences. Modern approaches to video generation have evolved from explicit latent modeling methods (such as VAEs and GANs) toward hierarchical stochastic processes (such as Diffusion and Flow Matching).

\paragraph{Notations}
To ensure consistency across formulations, we adopt a unified notation system.  
A video sequence is denoted as $\mathbf{x} = \{x_0, x_1, \dots, x_T\}$, where each frame $x_i \in \mathbb{R}^{H \times W \times C}$ represents a color image of height $H$, width $W$, and channel dimension $C$.  
Latent variables are denoted as $\mathbf{z} \in \mathbb{R}^d$, sampled from a prior distribution $p(\mathbf{z})$, while the true video data follow $p(\mathbf{x})$.  
Encoding and decoding processes are parameterized by neural networks $q_\phi(\mathbf{z}|\mathbf{x})$ and $p_\theta(\mathbf{x}|\mathbf{z})$, respectively, with $\phi$ and $\theta$ representing trainable parameters.  
For stochastic diffusion-based models, random noise at time step $t$ is written as $\boldsymbol{\epsilon}_t \sim \mathcal{N}(\mathbf{0}, \mathbf{I})$.  
\paragraph{Variational and Adversarial Latent Models}

Early generative video systems were grounded in latent-variable modeling, where the objective is to approximate the underlying data distribution by marginalizing over unobserved factors.  
A Variational Autoencoder (VAE) formalizes this through a probabilistic inference framework, optimizing the evidence lower bound (ELBO):
\[
\mathcal{L}_\text{VAE} = 
\mathbb{E}_{q_\phi(\mathbf{z}|\mathbf{x})}
[\log p_\theta(\mathbf{x}|\mathbf{z})]
- \mathrm{KL}\!\left(q_\phi(\mathbf{z}|\mathbf{x}) \,\|\, p(\mathbf{z})\right),
\]
which encourages faithful reconstruction while regularizing the latent distribution.  
When extended to videos, VAEs typically adopt 3D convolutional or recurrent encoders to maintain temporal coherence across frames.  
Although statistically grounded, VAEs often produce over-smoothed results due to the Gaussian assumption in $p_\theta(\mathbf{x}|\mathbf{z})$.

To enhance perceptual fidelity, Generative Adversarial Networks (GANs) replace explicit likelihood modeling with adversarial distribution matching.  
The generator $G_\theta$ and discriminator $D_\psi$ engage in a minimax game:
\[
\min_{\theta} \max_{\psi} 
\mathbb{E}_{\mathbf{x}\sim p(\mathbf{x})}\!\left[\log D_\psi(\mathbf{x})\right]
+
\mathbb{E}_{\mathbf{z}\sim p(\mathbf{z})}\!\left[\log(1 - D_\psi(G_\theta(\mathbf{z})))\right],
\]
where $G_\theta$ maps latent noise to samples and $D_\psi$ estimates their realism.
This adversarial formulation implicitly minimizes the Jensen–Shannon divergence between $p(\mathbf{x})$ and $p_G(\mathbf{x})$.  
For video generation, spatio–temporal discriminators and temporal-consistency losses are commonly employed to enforce coherent motion and continuity across frames.  
GAN-based methods deliver sharper results but remain sensitive to training instability and mode collapse, motivating probabilistic alternatives.

\paragraph{Diffusion and Flow Matching Models}

Diffusion models reinterpret generative learning as a denoising process that reverses a gradual corruption of data with Gaussian noise.  
Given a clean sample $\mathbf{x}_0 \sim p(\mathbf{x})$, the forward process produces noisy versions according to
\[
q(\mathbf{x}_t | \mathbf{x}_0) = 
\mathcal{N}\!\left(\sqrt{\bar{\alpha}_t}\,\mathbf{x}_0, (1-\bar{\alpha}_t)\mathbf{I}\right),
\]
where the noise schedule $\{\bar{\alpha}_t\}$ controls the signal-to-noise ratio.  
A neural denoiser $\boldsymbol{\epsilon}_\theta(\mathbf{x}_t,t)$ is trained to predict the noise, minimizing
\[
\mathcal{L}_\text{DM} = 
\mathbb{E}_{\mathbf{x}_0,t,\boldsymbol{\epsilon}}
\big[\| \boldsymbol{\epsilon}_\theta(\mathbf{x}_t,t) - \boldsymbol{\epsilon}\|_2^2\big].
\]
Sampling reverses this process, progressively refining random noise into structured video content.  
In practice, temporal conditioning and cross-frame attention are integrated to extend diffusion from static images to videos.

Flow Matching provides a continuous-time reformulation of diffusion through deterministic transport equations.  
By parameterizing the instantaneous velocity $\mathbf{v}_\theta(\mathbf{x}_t,t)$ between noise and data distributions, it minimizes
\[
\mathcal{L}_\text{FM} =
\mathbb{E}_{\mathbf{x}_0,t,\boldsymbol{\epsilon}}
\big[\|\mathbf{v}_\theta(\mathbf{x}_t,t) -
\mathbf{v}_t\|_2^2\big],
\]
where $\mathbf{v}_t=\frac{d\alpha_t}{dt}\mathbf{x}_0+\frac{d\sigma_t}{dt}\epsilon$ defines the interpolation path.  
This formulation supports deterministic sampling and enables efficient distillation of large diffusion models into compact one-step generators.  
Together, diffusion and flow-matching paradigms provide a physically consistent, likelihood-based foundation for scalable video generation, forming the backbone of generative augmentation in GAR.

\subsection{Autoregressive Models for Streaming Video Generation}
\label{sec:ar}

Converting traditional video generation models into a streaming generation paradigm is a crucial foundation for GAR. 
Classical autoregressive video generation models, such as CogVideo\citep{hong2022cogvideo}, VideoPoet\citep{kondratyuk2023videopoet}, Emu3~\citep{wang2024emu3} and MAGI~\citep{zhou2025taming}, decompose videos into spatiotemporal patches and map them into discrete tokens, generating videos token by token under the LLM-style autoregressive paradigm.
Due to the additional discretization process, mainstream autoregressive video generation approaches have shifted into a hybrid paradigm, performing diffusion/flow-based generation within frames while maintaining autoregression across frames. This line of work can be categorized based on the conditioning dependencies used for generating the current frame.

\paragraph{Teacher and Diffusion Forcing}
Pioneer models~\citep{gao2024vid} adopted the Teacher Forcing (TF) strategy~\citep{williams1989learning}, training next-frame prediction objectives conditioned on the ground-truth former context frames. 
Under this strategy, the video model depends on its previously generated frames during inference, but on ground-truth frames during training, leading to a growing discrepancy that causes performance to deteriorate rapidly as the sequence length increases, this phenomenon is also referred to as exposure bias.
To alleviate this, Diffusion Forcing (DF)~\citep{chen2024diffusion} introduced stochastic perturbations across temporal contexts.  Causvid~\citep{yin2025slow} and MAG-1~\citep{teng2025magi} exemplifies this recipe in streaming video generation, 
by assigning each frame an independent noise level and jointly denoising the entire sequence.
These methods expose the model to corrupted histories, then narrows the train–test gap and stabilizes long rollouts. 


\paragraph{Self Forcing}
To bridge the training–inference gap more fundamentally, Self-Forcing (SF)~\citep{huang2025self} reformulates autoregressive video diffusion as self-rollout: each frame is generated conditioned on the model’s own prior outputs, aligning training with inference and removing the mismatch left by TF/DF. Combined with few-step diffusion backbones and rolling KV caching, SF attains sub-second, real-time AR generation with stable long-horizon rollout. Self-Forcing++~\citep{cui2025self} further scales this recipe to minute-level sequences via stronger distribution matching and extended rollout training, while LongLive~\citep{yang2025longlive} demonstrates real-time interactive streaming with KV re-cache and frame-anchoring mechanisms that preserve global consistency during prompt switches.

Although Self-Forcing (SF) has gained widespread recognition in the field of autoregressive video generation, it still has limitations. For instance, it requires incorporating the inference process during training, which reduces training efficiency. Moreover, 
SF alone struggles to handle the challenges of infinite-length video generation, complex interactions, and dynamic scenes in GAR, thus necessitating the assistance of additional techniques, which we will introduce in the following subsections.

\subsection{Computational Efficiency Optimization for Video Generation Models}
\label{sec:eff}

Video generation models suffer substantial computational and memory demands from both multi-step sampling and the need to handle an additional temporal dimension. Recently, researchers have proposed several promising directions to improve the efficiency of video generation models, including advanced sampling, one-step generator, and feature/computation reuse strategies. 
%

\paragraph{Few-Step Solver}
Few-step solvers approximate the diffusion ODE/SDE solution by aggressively shortening the sampling trajectory. The DPM-Solver family~\citep{lu2022dpm,lu2025dpm,zheng2023dpm} gives an explicit ODE formulation that analytically integrates the linear term, typically cutting steps from hundreds to 5–20. PNDM~\citep{liupseudo} and DEIS~\citep{zhangfast} leverage multi-step history to compute the next update. UniPC~\citep{zhao2023unipc} unifies predictor–corrector updates for both ODE and SDE sampling, yielding stable, fast trajectories at very low NFEs without retraining. Learning-based discretization further tunes the schedule: AMED~\citep{zhou2024fast} uses a mean-value theorem view to estimate average velocity along the ODE path, while LD3~\citep{tonglearning} parameterizes time steps and learns an optimal few-step grid. Recently, SwiftVideo~\citep{sun2025swiftvideo} brings this recipe to video, aligning low-step with high-step trajectories to raise few-step video quality. Few-step solvers mitigate degradation at small step counts, though extreme settings (e.g., one–two steps) still face quality limits.


\paragraph{One-Step Generator} 
To overcome the quality limitations of few-step solvers under extremely few steps (e.g., one step), more methods aim to either modify the model training process or perform distillation on pre-trained generative models to achieve one-step generation.
For example, consistency training \citep{song2023consistency}, shortcut models \citep{fransone}, and Mean-Flow \citep{geng2025mean} learn the average velocity along the generative trajectory or shortcut paths to train a high-quality one-step generative model from scratch. On this basis, methods that distill pre-trained diffusion models into one-step models usually achieve better performance: Consistency Distillation \citep{song2023consistency} replaces the prior trajectories in consistency training with the pre-trained denoise trajectories, resulting in improved convergence and one-step generation quality; Progressive Distillation \citep{salimans2022progressive} gradually distills a multi-step generative model into a one-step model; and Distribution Matching Distillation (DMD) \citep{yin2024dmd} aligns the teacher and student models at the distribution level. GANs have also been used to assist one-step training or distillation due to their inherently single-step nature. For example, the Consistency Trajectory Model \citep{kim2024consistency} and DMD2 \citep{yin2024dmd2} incorporate a GAN loss into their original objective functions, achieving improved one-step performance.

\paragraph{Feature/Computation Reuse}
Besides the total number of sampling steps, the computational cost of each step also directly affects the overall efficiency.
Visual signals, especially videos, are generally considered to possess high redundancy \citep{he2022masked,tong2022videomae}. Meanwhile, the features between adjacent steps also exhibit certain redundancy \citep{ma2024deepcache}.
These observations have motivated techniques that improve the efficiency of diffusion/flow-based generative models by reusing local visual features (e.g. visual tokens) or intermediate computation results during sampling.
ToMeSD \citep{bolya2023tokendiff} first leverages the similarity between visual tokens to aggregate tokens in Stable Diffusion, improving generation efficiency with only a minor performance drop.
AT-EDM \citep{wang2024attention} further directly pruned unimportant tokens during the generation process.
Subsequently, SDTM \citep{fang2025attend} proposed a structural token merging technique that adapts to different noise levels, while ToFu \citep{kim2024token} combined both pruning and merging mechanisms. Vidtome \citep{li2024vidtome} applies token merging in video generation, improving temporal consistency while enhancing efficiency. 
DeepCache~\citep{ma2024deepcache} shifts focus to exploring the redundancy of features across adjacent sampling steps and introduced a training-free acceleration algorithm that reuse the high-level features of the U-Net during denoising. FORA~\citep{selvaraju2024fora} extends feature caching to Transformer-based diffusion models (DiT) and periodically recomputes the cached results at fixed intervals. TeaCache~\citep{liu2025timestep} further dynamically determines the caching strategy for each timestep through learning.

Current methods for computational efficiency optimization are mainly focused on image generation, with only a few attempts in video generation, such as SwiftVideo~\citep{sun2025swiftvideo}, Self-Forcing \citep{huang2025self} (which applies DMD \citep{yin2024dmd}), and TeaCache~\citep{liu2025timestep}. The effectiveness of these approaches in video generation and subsequent GAR applications remains to be verified. 
Potential challenges include the possibility that reduce d computational load in open and dynamic GAR scenarios may lead to greater performance degradation, as well as the potential adaptation issues of applying these paradigms to autoregressive models.

\subsection{Infinite-Length Video Generation and Quality Optimization}
\label{sec:long}


The ability to generate infinitely long videos is one of the core factors ensuring the quality of the GAR experience. However, as video length increases, two key challenges emerge: first, dense self-attention scales quadratically with sequence length, making long-range dependency modeling computationally intractable; Second, as the video length increases, the accumulation of errors becomes unacceptable. 

\paragraph{Long-Term Video Quality Optimization}
Beyond computational overhead, infinite-length video generation also faces the challenge of accumulated errors, which become increasingly severe as the video length grows dramatically. Self-Forcing~\citep{huang2025self} attributes the amplification of accumulated errors to exposure bias between training data and generated data, and addresses this by conditioning the generation of subsequent frames on previously generated frames during training. Later, Self-Forcing++~\citep{cui2025self} handle these accumulated errors by injecting denoised long-video latent variables into the initial noise used in Self-Forcing training, effectively extending the generation horizon of Self-Forcing by multiple times. In parallel, Rolling Forcing~\citep{liu2025rolling} incorporates the attention sink \citep{xiaoefficient} to retain the key–value states of the first frame as a global context anchor, thereby enhancing long-term global consistency. Besides anchoring on the first frame, MemoryPack~\citep{wu2025pack} simultaneously utilizes retrieved long-term video context and adjacent frames for long temporal coherence and motion and pose fidelity. Orthogonal to rollout alignment, Diffusion Adversarial Post-Training (APT)~\citep{lin2025diffusion} pushes per-frame generation to a single forward pass via adversarial post-training after diffusion pretraining, achieving real-time 2s 1280×720@24 fps clips and strong 1024 px images while not a long-horizon method, complements the above by improving one-step fidelity/latency and curbing step-wise drift in streaming settings. Despite successfully extending the duration of video generation, current approaches toward infinite-length video generation still exhibit limitations in GAR scenarios. For instance, existing minute-long video generation methods typically maintain consistency only within relatively simple or low dynamic scenes. Generating videos of unlimited length in GAR environments, where scenes can change drastically, remains a significant challenge.

\paragraph{Context Compression and Sparsification}
To tackle the intractable computation issue, several works compress or selectively route context information to preserve salient temporal cues while controlling computational growth.
LoViC \citep{jiang2025lovic} propose a context compression framework that distills historical tokens into compact representations, achieving near-linear scalability without sacrificing temporal alignment. Similarly, 
FramePack \citep{zhang2025packing} compresses the computational bottleneck of long video generation down to the image diffusion level by packing input frames into a fixed context length, regardless of the actual video duration.
Beyond static compression, \citet{cai2025mixture} reformulate long-context generation as a retrieval problem via a Mixture of Contexts (MoC) module, which adaptively selects the most relevant video chunks, enabling efficient long-term dependency over minute-scale sequences.
Complementary to this, \citet{li2025radial} developed a temporally distance-aware attention sparsification strategy, achieving an $O(n \log n)$ sparse attention mechanism that effectively extends the upper limit of video length.



\subsection{Multimodal Interactive Control in Video Generation}
\label{sec:control}
Compared to traditional video generation paradigms, GAR scenarios require the generation process to respond to multiple types of control signals across different modalities to ensure an immersive user experience. We enumerate the potential forms of control in GAR scenarios.

\paragraph{Camera Control} 
In GAR, users explore the AR world from a first-person perspective, requiring the GAR model to interact with head poses and gaze directions of users and generate visual content from the correct camera view. 
Such requirements can be achieved by incorporating camera control into video generation, transforming the user’s viewpoint into camera parameters to guide the content generation process.
SynCamMaster~\citep{bai2024syncammaster} and ReCamMaster~\citep{bai2025recammaster} encode camera extrinsics (i.e. rotation and translation) as latent vectors for explicitly viewpoint movement control.
To better capture the geometric relationship between the camera and the scene, Plücker embeddings \citep{sitzmann2021light} encode 3D lines as 6-dimensional vectors, providing a geometric representation for each pixel in a video frame. This embedding offers a comprehensive and compact description of camera pose information, then adopted in following cameral control based video generation, such as Cami2v \citep{zheng2024cami2v}, CameraCtrl~\citep{he2025cameractrl}, and Realcam-i2v \citep{zheng2024cami2v}

\paragraph{Drag Control} 
Drag-based interaction is an intuitive and effective way of controlling visual content, naturally possessing great potential for application in GAR interactions, such as gestures and touch.
Representative methods such as DragAnything~\citep{wu2024draganything}, Sg-i2v~\citep{namekata2024sg-i2v}, and DragNUWA~\citep{yin2023dragnuwa} generate motion trajectories through user drag operations, guiding content to move along specified paths. DragVideo~\citep{deng2024dragvideo} introduces a drag-based video editing framework, allowing users to manipulate local regions and control their motion. Building upon this, DragStream~\citep{zhou2025dragstream} enables real-time interactive video generation, where users can dynamically adjust and edit video content through drag operations during the generation process.

\paragraph{Audio Control}
GAR scenarios combined with ambient sounds or conversations can provide a more immersive interactive experience. This requires video generation models to have the capability to generate content consistent with the audio.
Incorporating audio control with video generation typically combining extracted multimodal audio features (e.g. Mel-spectrograms, rhythm, energy patterns, or semantic embeddings) with temporal feature vectors to synchronize visual content with corresponding sound events. For example, DreamTalk~\citep{ma2023dreamtalk}, SayAnything~\citep{ma2025sayanything}, and ACTalker~\citep{hong2025audio} leverage speech audio to generate corresponding head and lip movements. DabFusion~\citep{wang2025dance} and X-Dancer~\citep{chen2025xdancer} demonstrate audio-conditioned human motion generation in dance scenes with wide-ranging movements. 
\citep{zhang2024audio} and \citep{haji2024avlink} explore audio-driven video generation in more diverse everyday scenarios, such as the playing of various musical instruments, animal activities, and sports scenes.

\paragraph{Structure Control} 
Structural control incorporates displayed structural information, such as depth or skeletons, into the generative model to produce objects that adhere to given structural constraints. In GAR scenarios, this capability has the potential to accomplish complex gesture trajectory and pose skeleton controls, enabling more sophisticated interactions between the virtual and real worlds.
Pilot methods like ControlNet~\citep{zhang2023controlnet} or T2I-Adapter~\citep{mou2024t2iadapter} enable precise control over the spatial layout and structure of generated content by inject structural conditions into pretrained image diffusion. Subsequent attempts on video, for example, ControlVideo~\citep{zhao2025controlvideo} and SparseCtrl~\citep{guo2024sparsectrl} utilize depth maps for guidance, while SketchVideo~\citep{liu2025sketchvideo} and ToonCrafter~\citep{xing2024tooncrafter} take sketches as conditional input to achieve video generation or editing. 
More interesting attempts are reflected in human generation and animation, guiding human motion with skeletal maps, keypoint representations, or parametric human models (e.g., SMPL \citep{loper2023smpl}). These methods first extract human pose priors and then employ diffusion models to animate the reference character accordingly. Representative approaches such as Animate Anyone~\citep{hu2024animateanyone}, Human4DiT~\citep{shao2024human4dit}, and HumanDiT~\citep{gan2025humandit} use pose information to control human motion, while methods like LivePortrait~\citep{guo2024liveportrait} and HunyuanPortrait~\citep{xu2025hunyuanportrait} leverage facial landmarks to drive portrait animation. Such work provides a foundation for potential interactions with clearly structured objects in GAR scenarios, offering a promising avenue for ensuring interaction quality based on given structural priors.

In the field of video generation, there is extensive experience in exploring diverse controls. However, when applied to GAR, several challenges are anticipated. First, most existing generation control methods are based on traditional non-autoregressive video models, which limits the real-time interactivity between control signals and generated content. How to achieve high-quality, interactive, multimodal control within autoregressive GAR video models remains an open problem. Second, GAR scenarios often involve multiple control signals acting simultaneously, whereas current research primarily focuses on single-signal control. Consequently, enabling effective interaction under multiple simultaneous control signals is also a significant challenge.

\subsection{Scene and Asset Management}
\label{sec:scene}
GAR requires more sophisticated scene and asset management, primarily due to 
the requirement of scene and asset consistency under deeper integration of real–virtual interactions.


\paragraph{Explicit Management}
A straightforward way to ensure scene and asset consistency is to model them explicitly.
Such as ViewCrafter \citep{yu2024viewcrafter} , WonderJourney \citep{yu2024wonderjourney} and WonderWorld \citep{yu2025wonderworld} integrate an additional three-dimensional structure as the memory store. RoomTex \citep{wang2024roomtex}, SceneTex \citep{chen2024scenetex}, MeSS \cite{chen2025mess}, and Tex4D \cite{bao2024tex4d} achieve texture mapping on given scene geometries, resulting in complete and consistent scene generation. DiffusionRenderer \cite{liang2025diffusionrenderer}, Das \citep{gu2025das}, and UniRelight \citep{he2025unirelight} employ diffusion models as re-renderers to regenerate new videos based on explicit priors such as scene geometry and materials.

\paragraph{Implicit Management}
To more flexibly bypass the authoring barrier and enable generation without predefined modeling, GAR tends to rely more heavily on implicit representations of assets and scenes. Context-as-Memory~\citep{yu2025context} treats all historical frames as an external memory bank and retrieves relevant frames via camera field-of-view (FoV) overlap, enabling scene-consistent regeneration without explicit 3D reconstruction. WORLDMEM~\citep{xiao2025worldmem} extends this paradigm by storing both visual frames and state embeddings within a memory-attention framework, allowing accurate reconstruction of past environments and modeling their temporal evolution. On the multimodal front, LongVie~\citep{gao2025longvie} integrates dense (depth) and sparse (keypoint) control signals under unified noise initialization and global normalization, ensuring consistent motion and appearance across one-minute sequences. LongLive~\citep{yang2025longlive} brings these ideas into real-time interactive settings, introducing a KV-recache mechanism that refreshes cached attention states at prompt switches and a frame-sink token that maintains global anchors throughout streaming generation.

GAR abandon explicit representations such as meshes, materials, and textures in favor of implicit feature-based scene and content generation. While this approach lowers the authoring barrier and enhances generative flexibility and dynamism, it also introduces more severe consistency challenges, which will be one of the key issues for future GAR development.


%% file: secs/4_applications.tex
\begin{figure*}[!htbp]
    \centering
    \includegraphics[width=0.8\linewidth]{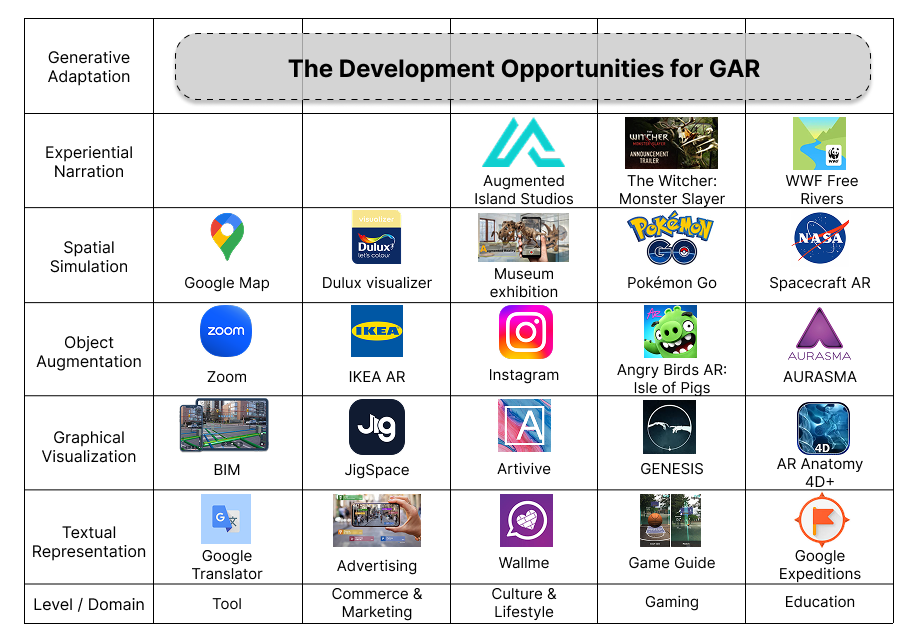}
    \caption{\textbf{AR Application Landscape.} The application landscape of Augmented Reality extends across five domains—Tool, Commerce, Lifestyle, Gaming, and Education—organized along a vertical axis of representational depth ranging from textual representation to generative adaptation.}
    \label{AR_Application_Landscape}
\end{figure*}

\section{Prospective Applications and Influences of GAR}
In this section, we survey how augmentation moves from current practice to emerging possibility and experiential shift. It first maps the application landscape~\ref{sec: AR-application-landscape}, then identifies new spaces and potentials~\ref{sec:Emergent-Application-Space-and-Potentials}, and finally examines transformations in experience, agency, and ecology~\ref{sec:GAR-Ecology}, laying the groundwork for later discussion of GAR’s broader influences.

\subsection{AR Application Landscape}\label{sec: AR-application-landscape}
Figure~\ref{AR_Application_Landscape} summarizes representative applications of AR across five major domains, including tool, commerce, lifestyle, gaming, and education. These domains reflect distinct purposes of AR adoption in practice, ranging from functional support in industrial settings to experiential engagement in cultural and learning scenarios.

\paragraph{Tool-oriented AR: From Representational Assistance to Collaboration.} In tool-oriented applications, AR serves as a cognitive extension that embeds digital information within the perceptual field of work and communication~\citep{allen2025workplace,10.1016/j.cag.2021.09.001, inbook}. Whether translating text, navigating through space, visualizing construction models, or framing telepresence, the underlying objective is to make abstract data actionable by reducing the distance between information and perception. Examples such as Google Translator AR, Google Map Live View, BIM-based (Building information modeling) visualization, and Zoom’s virtual backgrounds or avatars demonstrate this shared approach. 
These systems integrate linguistic, spatial, structural, and social information directly into the environment of use, turning observation into a form of reasoning~\citep{10138393, tahara2020retargetablearcontextawareaugmented}. Words appear where they are spoken, directions where one must walk, architectural plans where walls will rise, and visual personas where identities are performed. Across these different contexts, AR externalizes cognition by making thought perceptible and aligning perception with the immediate demands of a task~\citep{keil2020augmented,yang2019influences,10.5555/1894703.1894709}.

The strength of these systems lies in perceptual immediacy~\citep{10.1162/pres.1997.6.4.355}. By embedding information within the user’s field of action, AR shortens the loop between perception, decision, and execution, allowing users to reason within their environment rather than about it~\citep{10.1145/3706598.3714258,braker2023usercentered,yang2019influences}. A designer reading BIM overlays on site, a traveler following AR arrows through a crowded street, or a remote speaker adjusting self-presentation through virtual framing all participate in the same cognitive process, interpreting the world through a continuously updated visual scaffold. This perceptual integration enhances accuracy, situational awareness, and focus.

However, despite its immediacy, this augmentation remains reactive rather than generative. The system presents only what has been predefined, such as translations, paths, spatial plans, or visual backgrounds, and it cannot infer higher-level intentions or create alternatives when circumstances change~\citep{li2025satori,akcayir2017advantages}. The augmentation therefore functions as presentation rather than interpretation, providing information without the capacity for reasoning.

This limitation becomes evident whenever the surrounding context diverges from what the system expects. Translation overlays cannot adjust tone or register, navigation cues continue even when paths are obstructed, virtual backgrounds stay neutral despite emotional tension, and construction overlays fail to anticipate errors on site~\citep{HADI2024101842,10.1145/3743049.3743079}. These examples reveal a common structural constraint in current AR systems. They achieve precise alignment between digital and physical layers yet lack the ability to reason across them or generate new representations when the two fall out of sync.

\paragraph{Commerce and Marketing: From Visualization to Persuasion.} In commercial settings, AR operates as a persuasive medium that converts abstract advertising into perceptual experience~\citep{pozharliev2022effect,yang2020how,YIM201789}. Its main function is to reduce uncertainty by allowing users to see how products fit within their immediate surroundings. When consumers cannot physically examine an item, AR places it in their personal environment so that evaluation can occur through familiar cues such as light, scale, and texture. Applications including IKEA’s furniture preview, Dulux’s color visualizer, and Taobao’s virtual try-on demonstrate this approach. Each system reconstructs the product’s appearance within real space, aligning digital imagery with the user’s spatial and sensory perception.

The same mechanism operates in commercial presentation and industrial communication, where AR transforms explanation into embodied understanding. Platforms such as JigSpace enable companies to display mechanical systems or product architectures as interactive 3D models situated in real space.
Viewers can move around these models, examine individual layers, and activate animations that expose internal mechanisms, turning complex information into intuitive insight. Despite this immersive quality, such presentations remain limited by predefined scripts and fixed assets, allowing observation but not genuine participation.

The persuasive value of these systems lies not only in visual realism but also in perceptual coherence, which refers to the consistency between digital and physical cues within a unified environment. By linking simulated perception to everyday sensory expectations, AR transforms imagined utility into experienced utility~\citep{botinestean2025exploring,petit2022consumer}. Users gain confidence through embodied verification rather than rhetorical persuasion, following a cognitive process comparable to Bayesian updating, where each sensory response gradually refines prior beliefs about a product’s suitability or quality. However, this form of validation remains restricted to static simulation. The rendered scene is realistic but unresponsive, unable to adapt to changes in intention, lighting, or aesthetic mood. A system such as IKEA Place~\citep{ikea_place} can demonstrate how a chair fits within a room, yet it cannot infer whether the user prefers contrast or harmony, comfort or minimalism. Perceptual accuracy therefore, does not guarantee experiential relevance, since visual precision alone cannot generate emotional conviction or creative engagement.

\paragraph{Culture and Lifestyle: From Embodied Mediation to Experience.} AR has become a central medium for contemporary cultural and lifestyle communication by transforming spectatorship into participation~\citep{ramtohul2024augmented,chen2024why}. Across social platforms, creative industries, and cultural institutions, AR creates a shared perceptual environment in which visual symbols, narratives, and emotions are co-produced by humans and algorithms~\citep{dijkslag2024beautifya,szambolics2023adolescentsa}. The core mechanism operates through embodied mediation, meaning that cultural expression is translated into spatial, performative, and interactive forms that audiences can directly experience and modify~\citep{medinagalvis2024designing,swords2024emergence}. This process appears repeatedly in social media, live performance, film production, and museum interpretation, and together these practices illustrate a broader transformation in how cultural meaning is created and distributed~\citep{li2025satori,ramtohul2024augmented,petit2022consumer}.

On platforms such as TikTok and Instagram, AR effects operate as a shared visual grammar that turns gestures, moods, and facial expressions into recognizable cultural symbols~\citep{szambolics2023adolescents,li2023filters}. Filters and image effects allow users to externalize emotion and commentary through spatial imagery, merging affect and communication in a single expressive act. Yet these tools, which are predesigned and algorithmically optimized for rapid circulation, tend to generate aesthetic similarity rather than diversity. The creative freedom that AR appears to offer is therefore limited by its reliance on templates, as users modify existing visual patterns instead of inventing new ones.

A similar dynamic is visible in live-streaming and performance environments, where AR mediates between personal identity and public visibility. Virtual avatars, adaptive backgrounds, and environmental overlays help performers manage self-presentation, intimacy, and emotion while maintaining audience engagement~\citep{mills2025virtual,rahill2021effects}. However, these effects remain largely predetermined and cannot adapt to changes in tone or audience reaction. The outcome is a visual form that successfully conveys presence but lacks conversational responsiveness, creating performances that are enhanced by AR yet remain fixed in meaning.

Similar patterns appear in film production and museum design. In filmmaking, AR or even VR techniques enable directors and actors to perform within AR stages, where digital components are visualized at real scale to enhance spatial continuity and creative immersion. However, these systems remain limited by preauthored assets and fixed camera parameters, which allow visualization but not the generation of new narrative possibilities during production~\citep{10.1145/3706598.3714217}. A comparable condition exists in museum applications, including projects developed by the Smithsonian~\citep{smithsonian_ar} and MoMAR~\citep{momar}. These exhibitions transform cultural artifacts into interactive displays yet remain constrained by predetermined curatorial structures that provide the same interpretive sequence to all visitors regardless of interest or prior knowledge. In both domains, AR extends sensory engagement while leaving narrative adaptability largely unrealized.

\paragraph{Gaming and Entertainment: From Spatial Play to Narratives.} AR redefines the logic of play by shifting the game from a contained simulation to the continuity of lived space. Conventional video games create immersion by separating players from reality, while AR removes this separation and allows fiction to coexist with the physical world~\citep{article}. When virtual events take place within familiar surroundings, the player remains in the same environment yet perceives it as behaving differently. This overlapping of realities gives AR entertainment its distinctive aesthetic and presents immersion as a form of perceptual coexistence rather than withdrawal from the real world.

The central mechanism that enables this transformation is embodied participation. AR games link movement, gaze, and orientation directly to game states, turning the body into both controller and narrative tool~\citep{bektas2024gazeenabled,patibanda2023autopaizo}. In Pokémon Go, the act of walking through urban space determines encounter probability and spatial progress. The player’s physical movement becomes inseparable from exploration, and satisfaction arises not from symbolic control but from the rhythm of perception and motion. AR sports experiences apply similar principles, using full-body tracking to connect physical effort with digital response. The surrounding environment functions as a responsive stage that rewards agility, balance, and coordination. 
In all of these examples, the enjoyment of gaming from the integration of thought and action, as decisions are carried out through the same perceptual cycle that organizes ordinary movement.

From a design perspective, this embodied logic reshapes how space, agency, and narrative are organized. Titles such as The Witcher: Monster Slayer illustrate the integration of spatial movement with role-playing structure. Location data and camera perspective determine where and when story events unfold, requiring designers to anticipate shifting environments and brief attention cycles. The GENESIS AR card game follows a similar principle by combining physical cards with virtual effects, turning ordinary tables into active combat spaces. 
These hybrids reveal a dual condition in AR design, in which the real world supplies infinite variability while designers must still maintain coherence within that flow. Successful AR experiences depend on precise context calibration, which involves transforming environmental uncertainty into structured and meaningful interaction. The most engaging moments arise when digital features align with real-world conditions, creating a sense of spontaneous discovery that conventional virtual systems rarely achieve.

Players often describe this convergence as both empowering and unsettling~\citep{marto2022augmented,ku2021how}. Technical instabilities such as tracking errors, GPS (Global Positioning System) drift, and inconsistent lighting can easily disrupt the illusion of seamless integration within the gaming experience. Even successful titles such as Pokémon Go depend on event-driven participation and collective gatherings to sustain player engagement in the absence of a strong narrative thread. Consequently, current AR games tend to create brief episodes of immersion rather than continuous narrative involvement. The medium remains most effective at producing transient moments of wonder, while long-term narrative coherence continues to pose a challenge.

\paragraph{Education and Learning: From Visualization to Scaffolding.} In education, Augmented Reality redefines learning as a process that is both spatially embodied and perceptually grounded, transforming abstract concepts into interactive experiences that can be explored through movement and manipulation~\citep{mansour2025embodied,yang2025spatial,stalheim2024embodied,bos2022embodied,zhao2020augmented}. Applications such as Spacecraft AR~\citep{nasa_spacecraft_ar}, illustrate how AR makes hidden systems perceptible through visualization and interaction. Students can observe planetary motion within their classroom. Laboratory-based AR simulations extend this principle by enabling chemistry and physics students to conduct experiments that reveal molecular reactions and other invisible forces~\citep{LYRATH2023170, 10.1145/3688828.3699635}. The educational value of these systems lies in embodied cognition, in which knowledge is enacted rather than passively observed, allowing learners to build understanding through direct perceptual engagement.

These experiences also foster collaborative and inquiry-based learning. When groups of students examine the same AR representation of anatomy or ecology, they coordinate attention, gesture, and dialogue within a shared perceptual setting~\citep{10.1145/3700297.3700320}. The technology supports what researchers describe as joint embodied inquiry, in which participants work together through spatial exploration rather than through verbal explanation. Learners often refer to these situations as learning by seeing and doing, linking comprehension to the experience of a process instead of its memorization~\citep{10.1145/3742800.3742839}. In this way, AR transforms the classroom from an information space into a perceptual workspace where conceptual reasoning develops through embodied experimentation and collective interaction.

Despite these advantages, existing educational AR systems remain constrained in ways that limit their pedagogical depth, and these limitations could be mitigated through a generative design framework. One limitation concerns adaptivity. Most current AR lessons rely on linear scripts or fixed visualizations. Regardless of whether a student is struggling or excelling, the system presents the same sequence of activities, which makes it difficult to adjust challenge, pacing, or representation according to individual needs~\citep{akcayir2017advantages}. A second limitation involves instructional feedback. Although students are able to observe and explore, the system rarely interprets their gestures, gaze, or task performance to detect misunderstanding or redirect attention~\citep{zhang2025challenge}. As a result, teachers must interrupt the immersive process to intervene manually. A third limitation relates to scalability and flexibility of content. Creating a new AR lesson in fields such as biology, architecture, or social science requires specialized modeling and manual design, which prevents rapid development beyond a limited set of visually oriented topics~\citep{weerasinghe2022arigato,zekeik2025augmented}. Similar to issues in other domains mentioned above, these problems do not stem from the perceptual foundations of AR, but rather from static creation methods and insufficient contextual intelligence.

\subsection{Emergent Application Space and Potentials}\label{sec:Emergent-Application-Space-and-Potentials}
In Section~\ref{sec: AR-application-landscape}, we reveal the evolutionary trajectory of AR technology applications across various domains, progressing from perceptual enhancement to contextual intelligence. While current AR systems excel at aligning digital content with sensory reality, they remain constrained by fixed representations and non-generative logic. GAR promises to overcome these limits by enabling systems to not only display but also infer, adapt, and co-create meaning with users in situ. The following section explores this emerging design space.

\paragraph{Contextual Generation: From Static Overlays to Situated Reasoning.} In everyday uses of AR such as navigating a city, communicating abroad, or exploring unfamiliar spaces, users depend on the system’s ability to interpret their surroundings~\citep{10.1109/ISMAR.2008.4637362}. Yet conventional AR remains largely reactive. Navigation arrows continue even when routes are blocked, translation overlays ignore tone or formality, and visual annotations persist despite task changes. Such behaviors reveal a gap between spatial accuracy and situational understanding, as earlier studies have pointed out~\citep{10.1109/ISMAR.2008.4637360}. Speicher et al. described this misalignment as a failure of semantic adaptability~\citep{10.1145/3290605.3300767}, and Billinghurst and Dünser emphasized that most AR systems display rather than interpret~\citep{6171143}. The result is assistance that is technically precise but cognitively detached from users’ shifting goals.

GAR introduces contextual reasoning to bridge this gap. In navigation, generative models can translate sensor and language inputs into adaptive feedback by summarizing real-time traffic or weather conditions and rewriting guidance in conversational terms~\citep{10.1145/3706598.3713348}. For travelers, generative overlays can combine scene understanding with natural language to explain landmarks or adjust cultural tone, offering interpretations rather than literal translations. The same logic applies to accessibility contexts where information must adapt to sensory or emotional states~\citep{ubur2025augmentingcaptionsemotionalcues}. Visual–language models can identify fatigue or stress in the user’s behavior and simplify overlays or highlight safety cues accordingly. In these situations, generation functions as a dynamic mediator that reconstructs meaning in real time to align system feedback with user perception and intent.

Beyond text or instruction, contextual generation also operates through visual reasoning~\citep{nagy2025crossformatretrievalaugmentedgenerationxr}. When users move through complex spaces such as construction sites, museums, or city streets, the visible scene itself becomes an input for generative interpretation. Research in scene completion and neural reconstruction shows how AR systems can fill occluded regions, relight environments, or simulate future states of a space as the user changes perspective. Instead of freezing visual overlays, GAR can show how an environment might appear under different lighting, usage, or design conditions, helping users reason visually about transformation. In cultural or creative settings, generative relighting and style adaptation models such as IC-Light and SonifyAR extend this principle further, allowing AR to match the affective tone of a scene and shift between realistic and stylized renderings as the atmosphere changes~\citep{10.1145/3654777.3676406,zhang2025scaling}. These mechanisms translate context into perceptual continuity and ensure that visual augmentation evolves with both environment and emotion.

Across these emerging forms, contextual generation functions as an integrative intelligence that connects semantic and perceptual coherence. By combining scene understanding with generative synthesis, GAR allows the system to anticipate relevance, filter distractions, and reframe information as circumstances change. Whether the output is verbal guidance or visual adaptation, the underlying logic remains the same: perception informs generation, and generation refines perception. Everyday activities such as walking, reading, or observing become shared reasoning processes between user and system. This capacity for situated understanding establishes the foundation for more adaptive and participatory forms of mediation, where feedback evolves alongside experience rather than following a fixed script.

\paragraph{Adaptive Mediation: From Linear Scripts to Open Feedback Loops.} Many AR experiences today still follow linear scripts. A lesson proceeds through fixed steps, a museum tour plays a predetermined narration, and an industrial guide highlights components in a set order~\citep{6171143,10.1007/s10055-018-0347-2}. Such workflows exemplify what Azuma called the authorial model of AR, in which designers predefine how content unfolds while users simply receive it~\citep{inbook}. Although effective for structured instruction, these systems offer little flexibility when users’ pace, attention, or curiosity diverge from expectation. Once a learner pauses, a visitor lingers, or a performer improvises, the overlay continues regardless, producing assistance that is precise yet unresponsive.

GAR introduces adaptivity by embedding feedback mechanisms directly into the mediation process. In education, researchers such as Dunleavy and Dede have shown that immersive learning becomes most effective when systems interpret gestures and errors as part of a feedback loop~\citep{Dunleavy2014}. Building on this principle, generative models could enable AR lessons to vary explanation depth, visual emphasis, or pacing in real time. For example, an anatomy visualization might generate new analogies when a learner struggles with spatial reasoning or simulate simplified systems before gradually reintroducing complexity~\citep{10.1007/978-3-031-97763-3_30}. Rather than replaying scripted material, the system composes new representations that align with the learner’s current focus and progress.

Similar adaptive dynamics are likely to emerge in creative collaboration. Existing AR platforms such as Sketchar and Artivive already allow users to visualize sketches, paintings, and animations in physical space, bridging digital creativity with embodied experience. With generative assistance, these tools could evolve from display environments into collaborative partners that interpret brush strokes, predict composition intent, or suggest stylistic variations in response to gesture, color, and rhythm~\citep{10.1145/3746059.3747635}. Professional systems such as NVIDIA Omniverse XR or ARCore Scene Viewer may integrate generative feedback to refine spatial layouts and material options as teams exchange annotations or voice prompts. Through these mechanisms, GAR supports a new form of co-adaptive creation in which visual ideas are iteratively generated, tested, and reshaped in situ rather than planned in advance~\citep{10.1145/3654777.3676361}.
Communication and performance scenarios show the same potential for feedback-driven mediation~\citep{HADI2024101842}. In future hybrid meetings or live streaming contexts, generative overlays could analyze conversational sentiment and produce adaptive visual summaries that respond to audience attention. When a speaker emphasizes a key point, the system might generate a supporting diagram, and when confusion arises, it may replay or rephrase earlier content. In live art and entertainment, adaptive mediation will enable spontaneous interaction through generative lighting, effects, or virtual props that react to rhythm, gaze, and body movement, synchronizing with performers instead of following pre-rendered cues~\citep{10.1145/3613904.3642725}. In each case, the generative layer turns mediation from one-way delivery into an evolving dialogue among content, performer, and audience.

Across these domains, the core mechanism of GAR is co-adaptive control. By continuously sensing gaze, motion, and speech, the system constructs probabilistic models of engagement and regenerates content to sustain it. The interaction becomes cyclical as users act, the system interprets, and both adjust to maintain shared focus. As Qian et al. argue, feedback-rich environments redefine mediation as mutual regulation rather than presentation~\citep{10.1145/3491102.3517665}. Whether in classrooms, creative studios, or public performances, generative adaptivity allows AR to function as an interlocutor that negotiates rhythm, emphasis, and perspective in real time. This adaptive reciprocity forms the foundation for broader transformations in user experience and agency, as discussed in the following section.

\paragraph{Co-Creative Synthesis: From Template Editing to Generative Participation.} Contemporary social-media platforms have already turned AR into a participatory design medium. Snapchat’s Lens Studio and TikTok’s Effect House invite users to build and publish their own filters, enabling community-based circulation of visual effects~\citep{10.1145/3715275.3732126}. These creative workshops lower technical barriers and encourage viral experimentation, as filters and stickers become a shared visual language that spreads through social performance. Yet, this openness remains structurally constrained. Users assemble predefined assets and parameters within fixed interaction grammars. The result is a large volume of personalized yet stylistically convergent content, which the authors describe as aesthetic homogeneity~\citep{article}. Users may publish new filters, but the underlying logic of composition remains scripted and predetermined.

GAR introduces the possibility of authorship beyond templates. Rather than configuring presets, creators can communicate with the system through sketches, gestures, or natural language cues~\citep{MESSER2024100056}. In hybrid drawing applications such as Sketchar or mixed-media platforms like Artivive, generative models could interpret brush strokes or spatial arrangements to propose complementary textures, lighting, or narrative elements. A designer sketching in AR might ask the system to extend a line into a metallic surface or reimagine a mural under night lighting, receiving context-specific visual synthesis anchored in the real environment~\citep{zang2025airwearpersonalized3d}. The process becomes dialogic as the artist expresses intent, the system generates alternatives, and both iterate within the same perceptual space.

This generative participation is expected to extend to collaborative and performative domains. In shared AR studios, multiple participants could co-create installations that respond to collective motion or sound, with generative systems maintaining visual coherence as contributions overlap. Social storytelling and live streaming may evolve in similar ways, where avatars or virtual props adjust style and emotion in response to audience input or performer tone, creating scenes that transform in real time. Even casual short-form video makers could benefit from co-creative synthesis, as the system anticipates rhythm and mood to adapt typography, color palette, or camera movement dynamically~\citep{10.1145/3706598.3713417,Anderson_2025}. Across these examples, creativity shifts from editing static templates to orchestrating generative relations among people, space, and system.

This transformation aligns with what Manovich describes as the move from remix culture to co-evolutionary creation. GAR systems learn continuously from user behavior and aesthetic preference, producing contextual variations that in turn reshape those preferences. As boundaries between designer, audience, and algorithm begin to blur, AR becomes a living medium of negotiation that generates both content and the conditions for creativity itself. This co-creative synthesis broadens expressive range and establishes the conceptual bridge to the next section, which examines the experiential, agentic, and ecological implications of such generative partnerships.

\subsection{Transformations on Experience, Agency, and Ecology}\label{sec:GAR-Ecology}
The preceding analyses illustrate how GAR evolves from reactive assistance to adaptive and co-creative mediation. These advances no longer merely extend functionality but begin to reshape the nature of experience itself—altering how users perceive, act, and coexist within augmented environments.

\paragraph{Experiential Transformation: From Observation to Co-presence.} Most current AR systems still frame experience as an act of observation. Users look through a device to inspect, measure, or visualize additional information overlaid on reality~\citep{billinghurst2015survey}. Studies of AR usability consistently describe this mode as a perceptual extension, enhancing what users see without changing how they inhabit the scene. Even in immersive settings such as museum guides or educational simulations, interaction remains instrumental and externally directed\citep{10.1145/3607822.3614528}. The user engages with content but seldom feels that the environment itself is aware of their presence.

GAR begins to alter this dynamic by introducing responsiveness and continuity into perception. When the system interprets gaze, movement, or affective cues to regenerate what is seen, the augmentation no longer appears as an external layer but as a co-present entity that reacts to perception itself. For example, research on affective and adaptive AR shows how generative visual layers can synchronize with users’ emotional states by brightening tone, adjusting composition, or softening lighting to sustain attention or empathy~\citep{ubur2025augmentingcaptionsemotionalcues}. In artistic and narrative contexts, dynamic story environments integrate generative content that shifts mood and framing in response to user behavior, transforming observation into participation within a living scene~\citep{doh2025exploratorystudymultimodalgenerative}.

This experiential continuity extends to social and environmental presence. In collaborative AR, generative systems may adjust shared visual layers according to group proximity or conversational rhythm, creating what Billinghurst and Dünser described as shared perceptual fields~\citep{billinghurst2015survey}. The scene itself becomes a participant, mediating awareness among users. In urban or ambient installations, GAR can embed evolving imagery that responds to pedestrian flow, weather, or sound, allowing the environment to appear perceptually alive. Such experiences merge sensory feedback with contextual reasoning, leading users to inhabit rather than merely observe augmented space.

Across these developments, AR experience shifts from perceiving added information to co-existing with a generative presence. The user’s attention, emotion, and action become part of an ongoing perceptual negotiation in which the system contributes variation, rhythm, and atmosphere. Instead of functioning as a transparent interface, GAR operates as a responsive milieu that continuously re-authors reality and sustains a sense of mutual presence between human and environment.

\paragraph{Agency and Intentionality: From Interaction to Co-action.} In conventional AR, agency resides almost entirely with the user. Interaction is defined as a sequence of commands such as tapping to reveal layers, pinching to scale, or selecting predefined views~\citep{10.1145/2897826.2927365}. The system executes but does not participate. As Suchman observed in her study of human–machine planning, this model assumes a clear separation between human intention and computational response~\citep{10.5555/38407}. The user decides, and the system obeys. While this structure works for task-oriented visualization, it restricts how AR can support open-ended or evolving activities where goals change over time.

GAR introduces the possibility of shared intentionality. By linking perception, inference, and generation, the system begins to act as a co-agent that can propose, anticipate, and adapt actions. In design and creative workflows, early experiments such as SketchAR and Adobe Firefly AR prototypes demonstrate how generative assistants can suggest alternative compositions, predict spatial balance, or adjust camera framing as users move~\citep{10.1145/3563657.3596014}. Instead of following direct commands, the system interprets gestures and gaze as expressions of changing intent, generating complementary visual elements in response. Similar dynamics appear in motion-based domains. In mixed-reality performance environments, generative lighting and camera paths respond to performers’ gestures, creating an improvisational partnership where control circulates between human and machine~\citep{technologies12110216}.

This co-action transforms agency from individual control into distributed coordination. Users and systems negotiate outcomes through feedback and anticipation rather than execution alone. Studies in adaptive robotics and embodied AI describe this process as mutual regulation, where agents adjust their behavior to maintain a shared trajectory of meaning or goal pursuit. In AR contexts, this may appear as a navigation assistant that reinterprets environmental cues to match user behavior, or as a collaborative modeling tool that generates alternatives when the designer hesitates~\citep{10.1145/3613905.3643978}. The user’s gestures and pauses become cues for the system to act, while the system’s generative outputs invite reinterpretation and further adjustment.

Through these interactions, agency becomes fluid and negotiated. GAR systems do not replace user intention but extend it within a network of responsive possibilities~\citep{10.1145/3716553.3750776}. Each participant, whether human, algorithmic, or environmental, contributes partial intentionalities that converge through continuous regeneration. The outcome is not the result of command and execution but of dialogue and adaptation~\citep{10.1145/3706598.3713126}. In this co-active paradigm, the meaning of interaction expands to include the system’s interpretive initiative, redefining authorship and responsibility within augmented environments.

\paragraph{Ecological Shifts: From System Use to Hybrid Milieus.} As AR becomes embedded in everyday environments, its function extends beyond individual interaction to the orchestration of hybrid ecologies that interconnect humans, algorithms, and material systems~\citep{10.1145/3679318.3685408}. Traditional AR assumes discrete sessions of use, in which the user activates a device, completes a task, and then exits the augmented layer. GAR instead produces content that evolves with environmental and social inputs, remaining active across time and context~\citep{10.1109/TVCG.2008.24,BIMPAS2024110156}. The system no longer resides within a screen but diffuses into architectural surfaces, public spaces, and shared infrastructures, transforming augmentation from a tool into an ambient condition~\citep{10.1145/3279778.3281459}.

Early studies of situated media and urban AR installations already indicate this transition~\citep{10.1145/3441000.3441079}. Projects such as MoMAR, integrate environmental sensing with participatory visuals that change according to pedestrian flow, weather, or collective input. In these examples, augmentation functions not as a static overlay but as a generative ecosystem that mediates between physical and digital dynamics. As DiSalvo argues, such environments embody participatory materiality, where technological systems respond to environmental stimuli while shaping human patterns of movement and attention~\citep{10.1145/2556288.2557359}. Generative models further strengthen this reciprocity by synthesizing contextual imagery, narrative, or sound that evolves with usage patterns, allowing environments to learn and express.

These developments signal a shift from interaction to cohabitation. Within generative ecologies, users no longer operate isolated systems but inhabit spaces that continuously generate and adapt. In urban design, GAR can support context-aware facades or responsive signage that reconfigures according to collective behavior or environmental stress. In cultural contexts, public artworks may evolve through local data and visitor participation, forming what Giaccardi calls living heritage, a cultural artifact that reauthors itself through engagement~\citep{Giaccardi01052008}. Even within domestic settings, adaptive AR interfaces can coordinate lighting, temperature, and visual ambience based on inhabitants’ routines and preferences, merging computation with atmosphere.

GAR also gives technological form to these ideas by linking perception, reasoning, and environment in a continuous loop of sense-making. The system’s generative responses become part of the ecological feedback that shapes how people move, feel, and interact. As this feedback expands across urban, social, and cultural networks, the boundaries between media, infrastructure, and habitat begin to dissolve~\citep{inbook}.

In this emerging condition, AR is no longer an interface applied to the world but a medium through which the world itself becomes expressive and adaptive. The ecological consequence is twofold. Human experience gains new layers of responsiveness, while agency and authorship disperse into distributed systems whose behavior is only partially predictable. GAR therefore redefines the ecology of technology as an evolving milieu in which perception, computation, and culture develop together.

%% file: secs/6_conclusion.tex
\section{Conclusion}
The emergence of generative video modeling has expanded the scope of what can be rendered, perceived, and interacted with in real time. As generation moves from producing visual content to shaping interactive experiences, the boundary between media and environment begins to dissolve. Generative Augmented Reality (GAR) situates this transition within a coherent framework—where augmentation is achieved not by layering virtual objects, but by regenerating the perceptual world itself under the influence of sensing, intention, and interaction.

Rather than concluding a technological cycle, GAR opens one. It reframes augmentation as a living process that learns from engagement and composes reality through continual inference. The challenge ahead lies not only in scaling model efficiency and control, but in understanding how generative systems mediate presence, authorship, and meaning. As generative models become instruments of perception as much as of creation, augmentation evolves from an interface toward a mode of coexistence—where human and machine jointly construct what it means to see, act, and imagine within the same world.